\documentclass[apj]{emulateapj}
\usepackage{color}
\usepackage{multirow}

\newcommand{\dedx}{{\it dE/dx}}

\submitted{Accepted for publ. in ApJ, July 13, 2015}

\begin{document}
\title{
An Electron-Tracking Compton Telescope \\
for a Survey of the Deep Universe by MeV gamma-rays
}

\author{
T.~Tanimori\altaffilmark{1,2}$^{,*}$, 
H.~Kubo\altaffilmark{1}, 
A.~Takada\altaffilmark{1}, 
S.~Iwaki\altaffilmark{1},
S.~Komura\altaffilmark{1},
S.~Kurosawa\altaffilmark{3},
Y.~Matsuoka\altaffilmark{1},
K.~Miuchi\altaffilmark{4},
S.~Miyamoto\altaffilmark{1},
T.~Mizumoto\altaffilmark{1},
Y.~Mizumura\altaffilmark{2,1},
K.~Nakamura\altaffilmark{1},
S.~Nakamura\altaffilmark{1},
M.~Oda\altaffilmark{1},
J.~D.~Parker\altaffilmark{1},
T.~Sawano\altaffilmark{1},
S.~Sonoda\altaffilmark{1},
T.~Takemura\altaffilmark{1},
D.~Tomono\altaffilmark{1},
K.~Ueno\altaffilmark{1}
}

\affil{$^{1}$ Department of Physics, Kyoto University, Kitashirakawa-Oiwakecho, Sakyo-ku, Kyoto 606-8502, Japan}
\affil{$^{2}$ Unit of Synergetic Studies for Space, Kyoto University, Kitashirakawa-Oiwakecho, Sakyo-ku, Kyoto 606-8502, Japan}
\affil{$^{3}$ Institute of Materials Research, Tohoku University, 980-8577 Sendai, Japan}
\affil{$^{4}$ Department of Physics, Kobe University, 658-8501 Kobe, Japan}


\email{$^{*}: $tanimori@cr.scphys.kyoto-u.ac.jp}

\begin{abstract}
Photon imaging for MeV gammas has serious difficulties due to huge backgrounds and unclearness in images,
which are originated from incompleteness in determining the physical parameters of Compton scattering in detection,
e.g., lack of the directional information of the recoil electrons.
The recent major mission/instrument in the MeV band, {\it Compton Gamma Ray Observatory}/COMPTEL, which was Compton Camera (CC),
detected mere $\sim30$ persistent sources.  It is in stark contrast with $\sim$2000 sources in the GeV band.
Here we report the performance of an Electron-Tracking Compton Camera (ETCC),
and prove that it has a good potential to break through this stagnation in MeV gamma-ray astronomy.
The ETCC provides all the parameters of Compton-scattering by measuring 3D recoil electron tracks;
then the Scatter Plane Deviation (SPD) lost in CCs is recovered.
The energy loss rate (\dedx), which CCs cannot measure, is also obtained, and is found to be indeed helpful to reduce
the background under conditions similar to space.
Accordingly the significance in gamma detection is improved severalfold.
On the other hand, SPD is essential to determine the point-spread function (PSF) quantitatively.
The SPD resolution is improved close to the theoretical limit for multiple scattering of recoil electrons.
With such a well-determined PSF, we demonstrate for the first time that it is possible to provide reliable sensitivity in Compton imaging
without utilizing an optimization algorithm.
As such, this study highlights the fundamental weak-points of CCs.
In contrast we demonstrate the possibility of ETCC reaching the sensitivity below $1\times10^{-12}$~erg~cm$^{-2}$~s$^{-1}$ at 1 MeV.
\end{abstract}


\section{Introduction}\label{sec:introduction}
MeV gamma-ray astronomy provides the unique opportunity to study supernovae (SNe), as fresh radio isotopes in SNe emit MeV gamma-rays 
\citep{Matz_1988, Chevalier_1992, Iyudin_1994, Maeda_2012, Churazov_2014}.
Studies of active galactic nuclei and galaxies reveal the evolution of the early universe \citep{Zhang_2004, Inoue_2013}. 
Also, gamma-ray bursts are a promising probe to catch the first star \citep{Meszaros_2010, Nakauchi_2012}.
Especially, SNe are the most fascinating objects and are vigorously studied in all the fields  of astronomy.  Nevertheless,  there still remain the many fundamental  mysteries, such as, the origin of SNe type-Ia and nucleosynthesis. 
Although the thermonuclear explosion of a Single degenerate White Dwarf (SWD) has been believed to be the origin of SNe Ia 
and accordingly has been used as a distance standard in  cosmology, 
a merger of two WDs (DWD) has been frequently pointed out the more plausible origin  \citep{Hillebrandt_2000}.
Recently, the importance of the observations in MeV gammas to conclusively determine the origin was  remarked \citep{Summa_2013}, 
considering that  MeV gammas are the  unique probe that are directly emitted from the exploding or merging regions. 
They pointed out that a delayed peaking time (80 days, compared to  50 days for SWDs) appears in DWDs 
due to its higher total mass, in contrast to the  predicted model-independent peak time of 20 days from optical observations. 
Most of those sources are expected to be faint, because the number of the detected gammas is proportional to the cubic of the distance to the source.  Each detection will contain a substantial amount of uncertainties, such as, 
 fluctuations of $^{56}$Ni production($\pm$20\%) and viewing angles \citep{Maeda_2014}. 
We then estimate that detection of $\sim$100 SNe, each of which should have  $>$5$\sigma$ significance in 10 days of observation,  is required to obtain the statistically robust confirmation of these features.  
Given  the optical results  of $\sim$15 and $\sim$50/year SNe Ia and collapsars (SNe Ib, c, and II), respectively, within 60~Mpc \citep{Maoz_2012, Maeda_2014}, 
an MeV instrument with a sensitivity of $\sim$10$^{-12}$~erg~cm$^{-2}$~s$^{-1}$ for  10$^6$ s is needed to catch gammas of these SNe with the above-mentioned significance. 
Thus, to push such a new field in MeV-gamma astronomy, 
an instrument  with a high sensitivity of $\sim$1~mCrab ($3\times10^{-12}$~erg~cm$^{-2}$~s$^{-1}$ at 1~MeV in 10$^6$~s) is desired.

\begin{figure*}[htb]
\includegraphics[width=\linewidth]{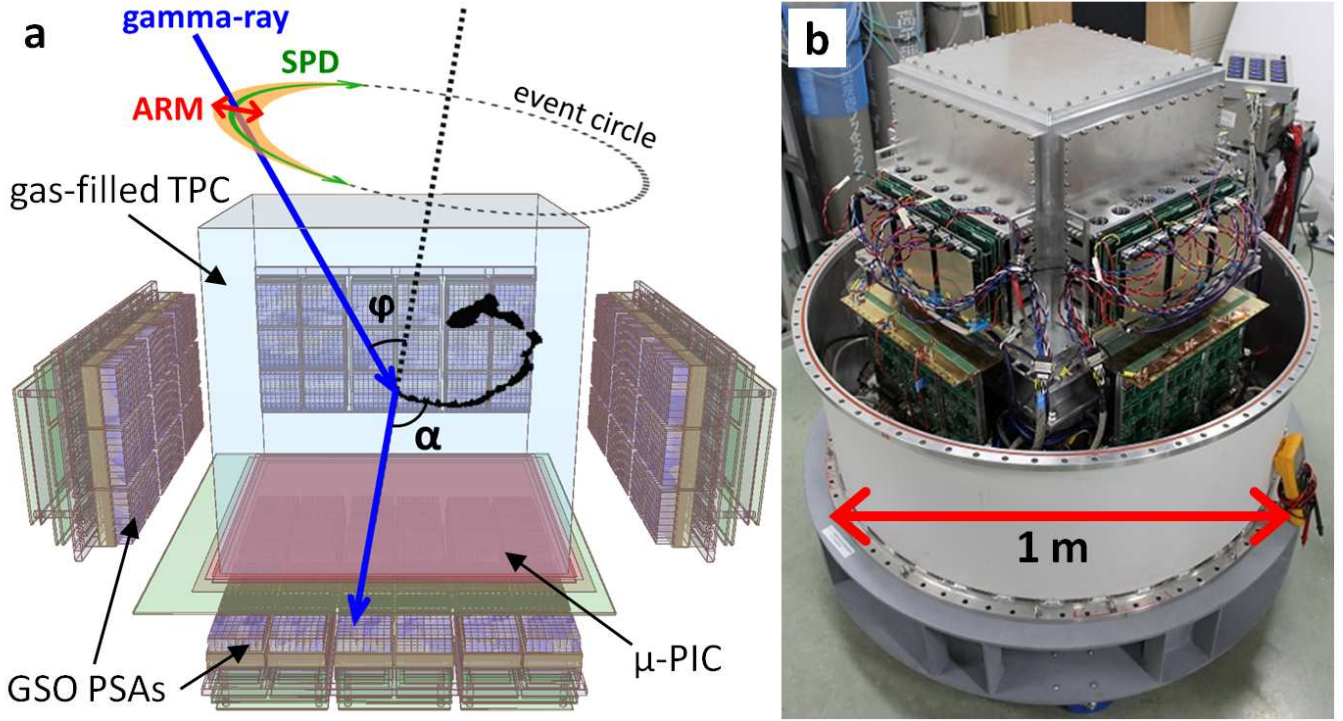}	
\vspace{1ex}
\caption{
(a) Schematic view of SMILE-II 30 cm-cubic ETCC. 
A micro-pattern gas detector (MPGD), which consists of 400 $\mu$m pitch pixels, is installed at  the bottom of the TPC, 
of which anodes and cathodes are connected via  strips to provide the 2-dimensional position and charge of the track. 
One PSA consists of 64 GSO bars (bar size: 6$\times$6$\times$13 mm$^3$) with 1 radiation length (R.L.). 
36 PSAs are put  at the bottom and 18 PSAs are on each side.
 A typical reconstructed track of a recoil electron is plotted in Fig.~\ref{fig:etcc_schematic}a, using an improved reconstruction method,
where the vertical width of the hit point represents the TOT as a pulse width of each pixel.
(b) Photograph of SMILE-II flight model instrument.
\label{fig:etcc_schematic}
}
\label{fig:etcc_schematic}
\end{figure*}

Historically,  COMPTEL operated with only $\sim$1/3 of the sensitivity expected from the calibration test before launch, and it may give a concern in estimating the true sensitivity in orbit.
The primary two causes of this discrepancy in COMPTEL are now understood as
 huge backgrounds in space and unclearness in the Compton images  \citep{Schonfelder_2000, Schonfelder_2004}.

After the close of the COMPTEL observations, it was pointed out that the  additional parameters, including  time of flight,
kinematical tests, and Scatter Plane Deviation (SPD) 
from the direction of the recoil electrons, would be necessary to reduce the background  for the next-generation  Compton cameras (CCs) \citep{Schonfelder_2004}. 
It is well known that in COMPTEL the time of flight between the forward and backward detectors dramatically reduced
the background and was a primary  factor for the success of COMPTEL.
Ideally,  the point spread function (PSF) of CCs must be evaluated, based on
 the two parameters of Angular Resolution Measure (ARM) and SPD, as shown in Fig.~\ref{fig:etcc_schematic}.
However, SPD is lost in CCs, due to  lack of direction information of the recoil electrons, thus the PSF shape is inevitably assumed to be rather spread (see Section~6). 
The recovery of the SPD by measuring  recoil-electron tracking is then  expected to both (1) improve 
the measured PSF by up to several degrees  
and (2) reduce a substantial amount of contamination of the  background leaked to the source region from the outside. 
In addition, the additional parameters of $\dedx$ in recoil-electron tracking are independent of  the reconstruction of Compton events, 
and therefore, use of them  would suppress the background dramatically without losing Compton events.

In contrast,  CCs with no additional parameter may lose a significant amount of Compton events in intense backgrounds, 
because  the application of cuts on physical parameters, such as, energy deposits or hit positions, 
produces a massive  uncertainty for the sensitivity. 
Nevertheless,  the recent trend in developing  advanced CCs concentrates on pursuing a larger  effective area and a  better energy resolution to improve the  ARM resolution. 
In fact, \citet{Aprile_2004} reported no detection of Crab in a balloon experiment with the liquid Xe CCs that has  the largest effective area of 20 cm$^2$  \citep{Aprile_2004}.
The NCT balloon experiment with the Ge-based CC and BGO shield detected  Crab with 4$\sigma$ \citep{Bandstra_2011},
although the reported number of signals from the Crab region was $\sim$1/6 of simulation over a large background. 
Recent  satellite proposals in MeV astronomy, in which the proposed sensitivities are larger than several mCrab, mention little on both of these problems \citep{Boggs_2006, Greiner_2009, Nakazawa_2012, Ballmoos_2012}. 
As such , it  seems to be difficult for even advanced CCs to reach the desirable  sensitivity in space.

\begin{table*}[tbp]
  \fontsize{6.5pt}{10pt}\selectfont
  \begin{center}
  \caption{
    Specifications of the instruments in Fig.~\ref{fig:effective_area}.
    \label{tab:config}
  }
    \begin{tabular}{lcccccc}
      \tableline \tableline 
      & TPC Size & Gas Parameters &Drift Velocity& Diffusion ($\mu$m/$\sqrt{\rm cm}$) & Anode-cathode & PSAs \\
      & (cm$^3$) & Mixture, Pressure, Drift Field &(cm~$\mu$s$^{-1}$)& Transverse/Longitudinal & Coincidence &  \\
      \tableline
      SMILE-I type prototype & 10$\times$10$\times$8 & Ar 90\%/C$_2$H$_6$ 10\%, &4.0& 470/230 & online & GSO \\
      & & 1 atm, 400 V~cm$^{-1}$ & & & 10~ns gate & 1~R.L. \\
      SMILE-I Flight Model & 10$\times$10$\times$15 & Xe 54\%/Ar 40\%/C$_2$H$_6$ 6\%, &2.4& 500/290 & online & GSO \\
      & & 1 atm, 380 V~cm$^{-1}$ & & & 10~ns gate & 1~R.L. \\
      SMILE-II small prototype & 7.5$\times$7.5$\times$15 & Ar 90\%/C$_2$H$_6$ 10\%, &3.6& 470/300 & offline & GSO \\
      & & 1 atm, 170 V~cm$^{-1}$ & & & adequate gate & 1~R.L. \\
      SMILE-II Flight Model & 30$\times$30$\times$30 & Ar 95\%/CF$_4$ 3\%/iso-C$_4$H$_{10}$ 2\%, &6.5& 300/300 & offline & GSO \\
      & & 1 atm, 160 V~cm$^{-1}$ & & & adequate gate & 1~R.L. \\
      SMILE-II (PSAs improved) & 30$\times$30$\times$30 & Ar 95\%/CF$_4$ 3\%/iso-C$_4$H$_{10}$ 2\%, &6.5& 300/300 & offline & GSO \\
      & & 1 atm, 160 V~cm$^{-1}$ & & & adequate gate & 3~R.L. \\
      SMILE-III & 30$\times$30$\times$30 & CF$_4$ 100\%, &5& 100/100 & offline & GSO \\
      & & 1 atm, 160 V~cm$^{-1}$ & & & adequate gate & 3~R.L. \\
      Satellite-ETCC (1 module) & 50$\times$50$\times$50 & CF$_4$ 100\%, &5& 100/100 & offline & LaBr$_3$ \\
      & & 1 atm, 160 V~cm$^{-1}$ & & & adequate gate & 10~R.L. \\
      \tableline
    \end{tabular}
    \tablecomments{All the drift velocities and diffusion constants were calculated using MAGBOLTZ simulation code \citep{Biagi_1999}.}
  \end{center}
\end{table*}

To achieve the desired  high sensitivity, combination of the sharp PSF and the reduction of the background with  the SPD 
and additional parameters from recoil-electron tracking is  a promising approach, given  a MeV instrument with the  effective area of several tens of cm$^2$ will reach  a sensitivity of $\sim$1 mCrab if  perfect background rejection is applied. 
Only a few  studies on CCs with recoil electron tracking have been reported  so far \citep{Kanbach_2006, Vetter_2011}.
Among them , SMILE-I (``Sub-MeV gamma ray Imaging Loaded-on-balloon Experiment'': first Electron-Tracking Compton Camera; ETCC) showed 
the possibility to remove most of the background without an active shield by
recoil electron tracking in a gaseous time projection chamber (TPC) \citep{Takada_2011a}.

Here we present the performance of an improved 30 cm-cubic ETCC for our second balloon experiment (SMILE-II). 
We have successfully made  robust reduction of background and obtained clear images  by 3D electron tracking. 
In Section~\ref{sec:etcc}, the concept and structure of the ETCC are concisely  explained.
Then  its fundamental performance is presented in Section~\ref{sec:performance}. 
The durability of the ETCC with  excellent ability of reducing background under intense radiation condition is mentioned in Section~\ref{sec:intense_bg}. 
In Section~\ref{sec:spd}, an improvement of imaging, thanks  to the use of SPD, is described in detail, 
and the definition of the PSF of the ETCC to reach the ultimate sensitivity of the CC is thoroughly discussed in Section~\ref{sec:discussion}. 
Finally, we summarize the characteristics  of the ETCC briefly and consider the  prospects for the future MeV gamma-ray astronomy with the use of advanced  ETCC.

\section{Electron-Tracking Compton Camera}\label{sec:etcc}
An ETCC consists of a TPC based on a micro pattern gas detector (MPGD) with 400~$\mu$m pitch pixels for 3D tracking of recoil electrons 
and Gd$_2$SiO$_5$:Ce (GSO) pixel scintillator arrays (PSAs) to measure  scattered gamma rays \citep{Tanimori_2004}.
With an ideal effective area of 100 cm$^2$ at 1 MeV for a 50 cm-cubic 3 atm CF$_4$ gas,
such a TPC would be a good device for a CC with electron-tracking.
CF$_4$ gas is an ideal gas for ETCC,
with significant benefits, including  a large cross section for Compton scattering,
small photo-absorption cross section, and small diffusion constant (see Section~\ref{sec:discussion} for detail). 
Several time-projection chambers  have already been operated with pure CF$_4$ in accelerators and underground experiments \citep{Thun_1988, Isobe_2003, Miuchi_2010}. 
However, we need more study to use the CF$_4$-dominant gas in the MPGD-based TPC, due to the requirement of a higher-voltage electric field on the MPGD. 
In particular, good  reliability for the stable operation of the TPC is crucial in the balloon experiments, 
and hence we at present use
Ar-based gases with some additional quencher gases
to study the effectiveness of electron tracking.
Table~\ref{tab:config} summarizes the  detailed components of the Ar-based gases used in our experiments.

3D tracking of the recoil electron provides  the incident gamma-ray direction as an arc by adding SPD to ARM (the  additional angle of $\alpha$ in Fig.~\ref{fig:etcc_schematic}a), 
and the energy-loss rate ($\dedx$) of the track,
all of which are effective tools to reduce backgrounds by directional cut, kinematical tests and particle identification, respectively. 
Note we  have already examined the concept of ETCC \citep{Orito_2004, Tanimori_2004, Takada_2005},
including an ARM resolution of 4$^\circ$ (FWHM) at 662 keV with LaBr$_3$ PSAs for medical imaging \citep{Kabuki_2010, Kurosawa_2010}.

The readout method of the TPC is a key technology of this balloon experiment, as described in \citet{Tanimori_2004} and \citet{Takada_2011a}. 
 Only one or a few tracks, including background for Compton events, is found to appear in the TPC.  
Given that, conventional orthogonal strip electrodes, instead of pixel electrodes, as used in  typical TPCs, could be used to reconstruct the 3D tracks 
by slicing along the direction of the electron drift using fast timing, 
and by combing the anode and cathode hits that have  the same times \citep{Tanimori_2004}.
Then,  we could reduce the number of readout electrodes of the TPC by an order of two  for this 30 cm-cubic ETCC, 
which is an allowable level using recent ASIC and FPGA, while satisfying the severe constraint of balloon experiments for power consumption, space and cost. 
However, such a drastic reduction of the number of signals would also introduce a risk of increasing  the ambiguity of electron tracking, as mentioned later.

\begin{figure*}[htb]
\includegraphics[width=\linewidth]{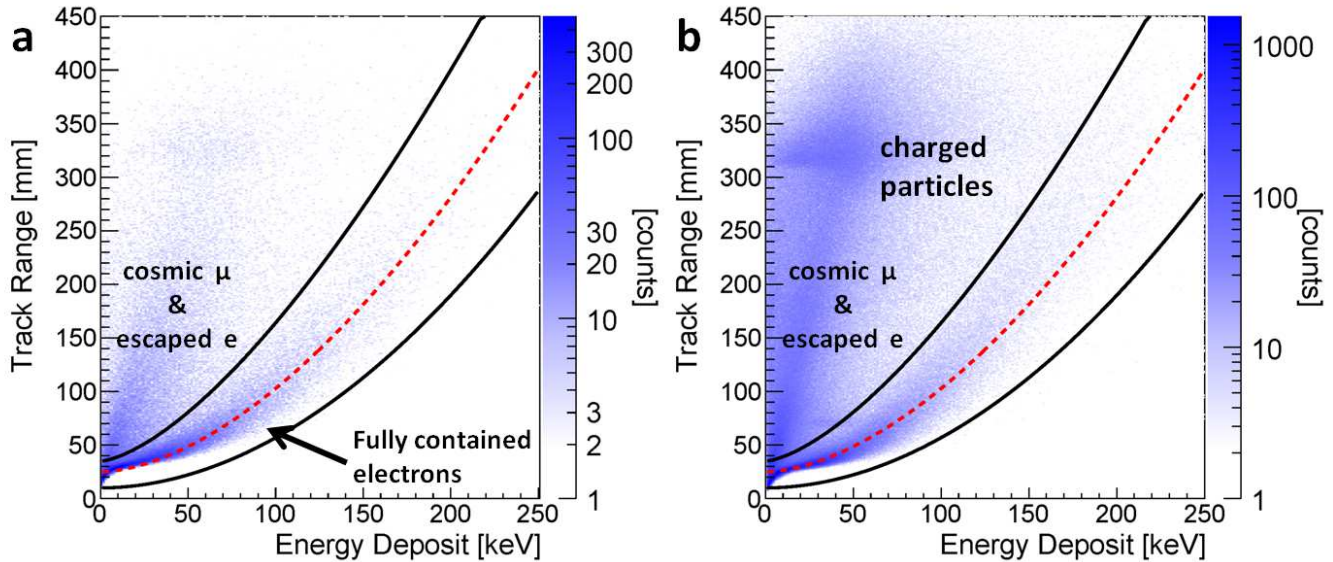}	
\vspace{1ex}
\caption{
Correlations of the track length and energy deposit in the TPC 
(a) under the condition of gamma-ray irradiation from $^{137}$Cs (3 MBq) at a distance of 1 m in the laboratory, 
(b) under the intense radiation generated by a 140 MeV proton beam. 
In all figures, the fully contained electrons are clearly separated from the minimum ionizing charged particles, 
e.g.,  cosmic muons, high-energy electrons, and neutron-recoil events.
\label{fig:dedx}
}
\end{figure*}

For MeV astronomy, SMILE-I with a 10 cm-cubic ETCC was performed in 2006.  
We selected 400 downward gamma-rays from 2.3$\times$10$^5$ triggered events, using the kinematical test and $\dedx$ \citep{Takada_2011a}, 
where a kinematical test was made with  comparison of the measured angle $\alpha$ to the calculated one,
$\Delta_{\alpha}$ as mentioned in Section~\ref{sec:etcc} of \citet{Takada_2011a}.
Although both the resolution for $\dedx$ and the detection efficiency were inefficient, due to a low tracking efficiency of $\sim$10\%,
in which at least 5 hit points on each track were required,
the combination of the $\dedx$ cut and kinematical test was found to be useful for the reduction of neutrons and cosmic rays.
 This inefficiency was not due to the low gain of the MPGD but to a bug in the algorithms set in the FPGA used in the readout electronics (see the next section for detail). 
The obtained diffuse atmospheric and cosmic fluxes were consistent with past observations. 
In addition, it is noted that its cubic structure provides a large field of view (FOV) of 2$\pi$ str.

\section{Basic Performance of the SMILE-II ETCC}\label{sec:performance}
We have been attempting to establish  a high sensitivity for the ETCC,  using a 30 cm-cubic ETCC (Fig.~\ref{fig:etcc_schematic}b) 
with a balloon experiment for the detection of the Crab (SMILE-II) \citep{Takada_2011b}. 
To detect  Crab with a reliable $\sim$5$\sigma$ significance in an exposure of  several hours, 
an effective area of 0.5 cm$^2$ at 300 keV and an ARM resolution of $<$10$^\circ$ (FWHM) at 662 keV are required  \citep{Tanimori_2012, Sawano_2014}.

The improvement of the reconstruction of tracks in the TPC was an essential factor for SMILE-II. 
For SMILE-I, only addresses of the anode and cathode strips hitting simultaneously within a 10~ns gate were encoded. 
However, a 10~ns gate is too short to get all hit strips, due to the delay in the encoding circuit. 
This was the reason for the small effective area of SMILE-I. 
To recover all hit points in SMILE-II, all addresses of the hit strips on anode and cathode with their hit timings are recorded without requiring the coincidence of 10~ns, 
and an adequate gate is applied in the analysis \citep{Tanimori_2012}. 
In addition, a pulse height of each strip was recorded as a timing width over the threshold (TOT).
Fig.~\ref{fig:etcc_schematic}a shows  a typical reconstructed track of a recoil electron for the TPC of SMILE-II,  using an improved reconstruction method,
where the vertical width of the hit point represents the TOT as a pulse width of each pixel.
To measure the precise energy deposit in the TPC, a sum of 64 strips of both cathodes and anodes are fed to Flash ADC (25 MHz), 
and its wave form was recorded in  10~$\mu$s. 
Also, two neighboring strips of both anodes and cathodes were combined to increase the pulse height. 
Then the track reconstruction efficiency was dramatically improved from $\sim$10\% to 100\%, 
which provides a much better resolution for $\dedx$, as demonstrated  in Fig.~\ref{fig:dedx}a \citep{Mizumura_2014},
compared to that of SMILE-I  (Fig.~7 in  \citet{Takada_2011a}). 
The detail of the performance of this TPC is described in \citet{Matsuoka_2015}. 

\begin{figure}[htb]
\includegraphics[width=\linewidth]{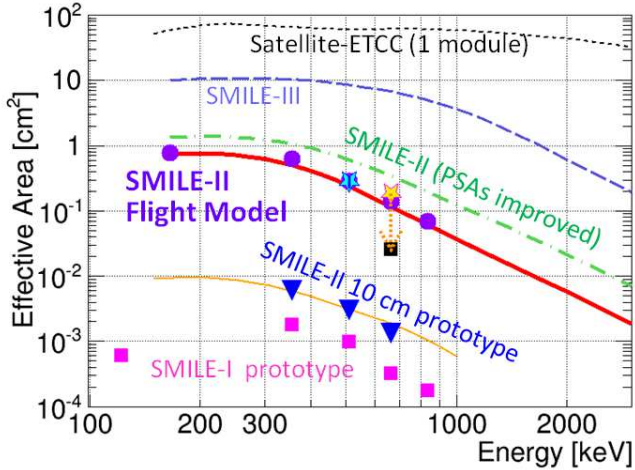}	
\vspace{1ex}
\caption{
Energy dependences of several measured and simulated effective areas drawn as points and lines, respectively.
SMILE-I prototype (magenta squares), SMILE-II small 10 cm-cubic ETCC prototype (blue inverted-triangle points and yellow line), 
SMILE-II 30 cm-cubic ETCC flight model (purple circles and red line), improved one with 3 R.L. GSO PSAs (green dotted-dashed line), 
a 30 cm-cubic ETCC of SMILE-III (blue dashed line), and a 50 cm-cubic ETCC (black dotted line). 
The effective areas were measured at several energy points, using RI sources of $^{139}$Ce, $^{133}$Ba, $^{22}$Na, $^{137}$Cs and $^{54}$Mn.
The detailed configurations of these ETCCs are described in Table~\ref{tab:config}. 
Additionally, the yellow and blue stars are the effective areas of
the Fig.~\ref{fig:RCNP}b beam experiment and Fig.~\ref{fig:weak_src}b Crab simulated measurement, respectively. 
The black square is explained in Fig.~\ref{fig:RCNP}d.
\label{fig:effective_area}
}
\end{figure}

Figure~\ref{fig:dedx}a shows the track range and its energy deposit for 662 keV gammas 
from $^{137}$Cs source set at a distance of 1 m from the ETCC, in which  $\dedx$ is a gradient of the distribution. 
The measured $\dedx$ in this figure clearly distinguishes the Compton electrons fully contained in the TPC from the backgrounds, 
and resultantly enables us to remove most of the backgrounds without loss of Compton events by applying cuts. 
The red-dashed line in this figure is the empirical formula of $\dedx$, which is approximately proportional to E$^{1.72}$ \citep{Sauli_1977}, 
for recoil electrons fully contained in the TPC. 
The region within E$^{1.72\pm0.22}$ contains $\sim$95\% of Compton events inside ($\dedx$ cut), as supported by the simulation result. 
In this article, all simulations were done using GEANT4 \citep{Agostinelli_2003}.
 The simulation also shows the distribution of higher energy particles, such as, electrons escaping from the TPC 
and minimum ionizing particles (cosmic muons) on the upper boundary, as  actually seen in Fig.~\ref{fig:dedx}a. 
In addition, the scattering point of the track is required to be within 1 mm inside the drift region of the TPC in order to remove events 
originated from  gammas that  scattered from the wall of the drift cage of the TPC (fiducial cut), with  which $\sim$10\% of
the events passing the $\dedx$ cut were removed. 
Thus, by applying only two simple cuts, almost all of the fully contained Compton events are obtained, after  several kinds of backgrounds are filtered out. 
For the obtained  fully contained Compton events, effective areas were measured at several energy points, 
using RI sources of $^{139}$Ce, $^{133}$Ba, $^{22}$Na, $^{137}$Cs and $^{54}$Mn, along with the simulation results for several types of ETCCs, 
where the effective area was obtained from the number of detected gammas 
within twice the FWHM of the energy resolution centered at  the gamma energy in  the background-subtracted energy spectrum. 
The simulation  results were obtained mainly from the production of two probabilities: 
that of Compton scattering and its recoil electron contained fully in the TPC, and that of the full absorption of scattered gammas in the PSAs, 
where the absorption of the materials of the TPC and the supporting frames are taken into account.

Figure~\ref{fig:effective_area} shows good consistency between the measured efficiencies for the ETCCs 
listed in Tab.~\ref{tab:config} (shown as points) and their simulation results (shown as lines). 
Here the 10 cm-cubic SMILE-II ETCC prototype (blue inverted-triangle points and yellow line), 
30 cm-cubic SMILE-II flight model ETCC (purple circles and red line), 
improved flight model with 3 radiation length (R.L.) GSO PSAs covering the bottom half of the TPC (green dotted-dashed line), 
and SMILE-III ETCC  (blue dashed line; described in the later section) are given.  
In addition, a 50 cm-cubic ETCC with CF$_4$ gas at 3 atm and 10 R.L. PSAs ($\Delta E/E\sim$4\% at 662 keV: FWHM) is plotted (black dotted line), 
in which the PSAs are set within the pressure vessel to catch high-energy electrons escaping the gas volume of the TPC. 
 A large FOV of 2$\pi$ str similar to SMILE-I was confirmed for the 30 cm-cubic ETCC \citep{Matsuoka_2015}. 
Thus, we are able to estimate precisely the effective area for the extensions of the volume, 
gas type and pressures of the TPC, or the R.L., and the better energy resolution of PSAs. 
These simple extensions ensure an effective area of 11 cm$^2$ at 300 keV for a following balloon experiment with a long duration flight (SMILE-III) 
and that of 60 cm$^2$ with 2$^\circ$ (ARM, FWHM) at 1 MeV for a 50 cm-cubic ETCC with CF$_4$ gas at 3 atm and $\sim$10 R.L. PSAs, e.g.,  LaBr$_3$. 
Furthermore, a satellite-ETCC consisting of four 50 cm-cubic ETCCs 
for a middle-class satellite  would reach ~240 cm$^2$ (4$\times$60 cm$^2$) with an ARM resolution of 2$^\circ$ at 1 MeV.

\begin{figure*}[htbp]
\includegraphics[width=\linewidth]{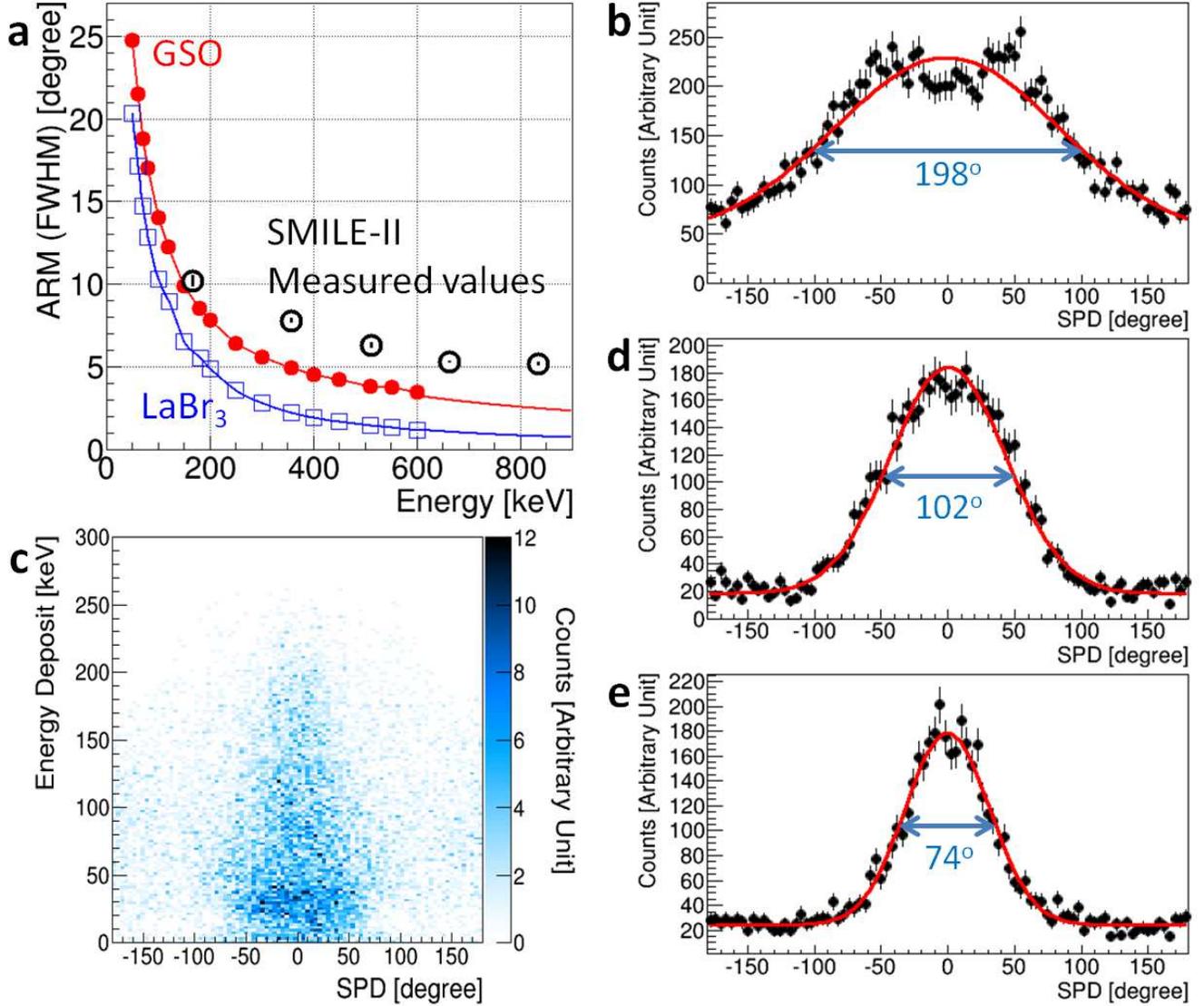}	
\vspace{1ex}
\caption{
(a) Variation of the measured ARM resolutions and those simulated for GSO and LaBr$_3$ PSAs are 
plotted with  open circles, filled circles and open boxes, respectively. 
Here statistical errors of the measured ARM resolution are indicated within the plots, 
and simulated ARM resolutions are  calculated from the energy resolutions of the TPC and PSAs. 
(b) The distribution of the SPD using the previous tracking method. 
(c) Correlation of the energy of a recoil electron and the SPD resolution after the improvement of the tracking method, 
and its projections on the SPD for 
(d) low energy recoil electrons (10-80 keV) and
(e) high energy ones ($>$80 keV). 
 All the figures include some amount of background from gammas scattered
between the $^{137}$Cs source and the ETCC due to the worse energy resolution of GSO PSAs, 
which broaden the SPD resolution, compared to that due to directly incoming gammas.
\label{fig:arm_spd}
}
\end{figure*}

 The ARM resolutions were obtained at the same energy points (Fig.~\ref{fig:arm_spd}a) 
with those calculated, based on  the detector energy resolution. 
 The measured ARMs were close to the theoretical limit of  the resolution expected for  the detector energy resolutions. 
Fig.~\ref{fig:arm_spd} also gives the  ARM resolutions for future ETCCs with  scintillators with better energy resolution. 
The discrepancies between the measured and calculated ARM resolutions are considered to arise from the $\sim$8 mm uncertainty 
in the track for the scattering point of the gamma, due to the worsened 3D reconstruction of the recoil electron, as mentioned below. 

\begin{figure*}[htb]
\includegraphics[width=\linewidth]{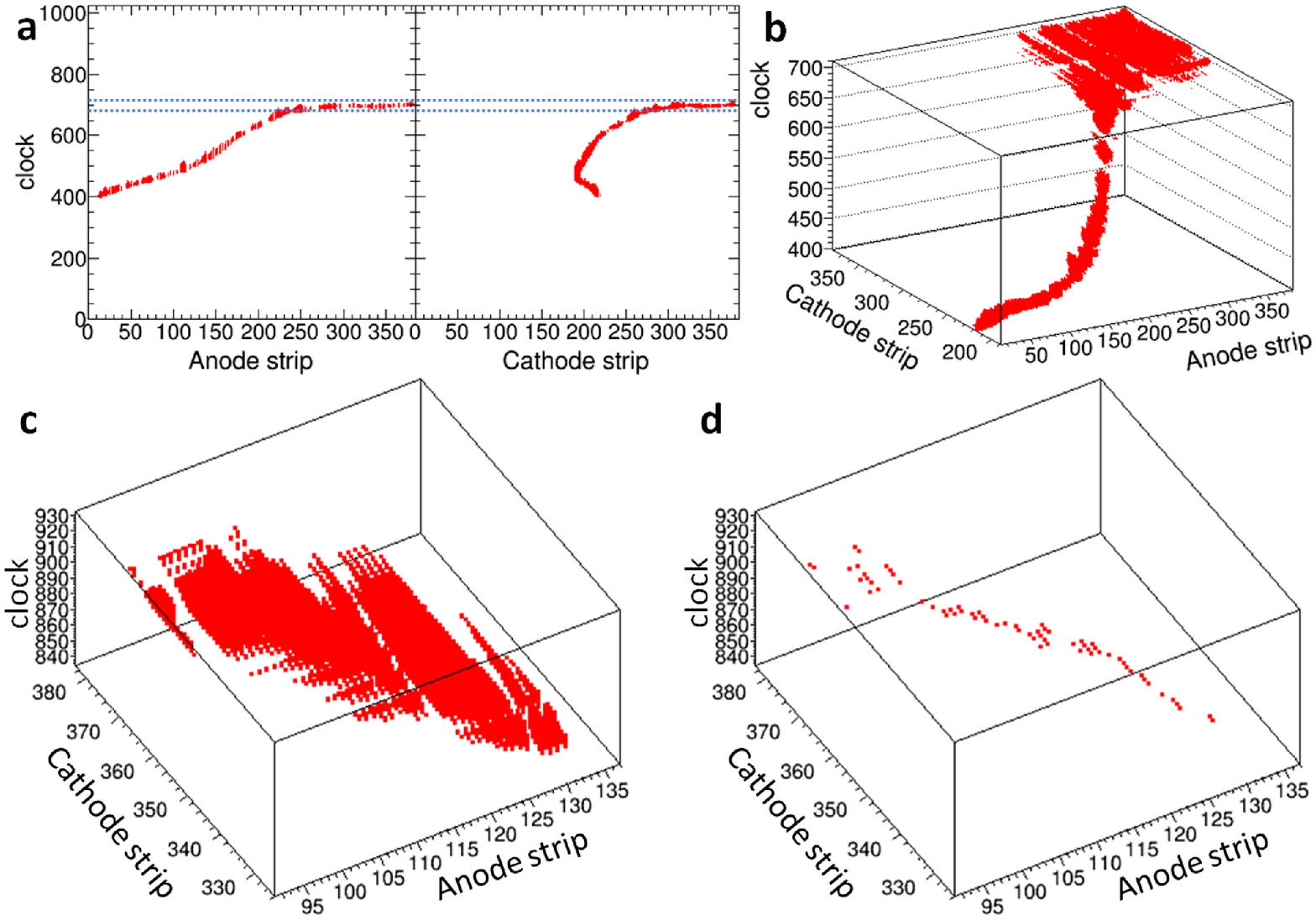}	
\vspace{1ex}
\caption{
(a)Schematic views of the track projected on the side wall of the TPC, 
and (b) the 3D electron-track images with the ambiguity from the multi-hits during 10 ns indicated 
between two dotted blue lines on the orthogonal 2D strip readout, 
and typical electron track (c) before and (d) after the correction of time walk.
\label{fig:tracks}
}
\end{figure*}

Finally, an SPD can be determined for all the events, and we obtained a resolution of SPD of 200$^\circ$ (FWHM)  (Fig.~\ref{fig:arm_spd}b), 
which is about two times worse than the expected SPD resolution, due to multiple scattering in Ar gas. 
The SPD is determined with  a linear fitting of the entire track. 
Even a SPD with such a poor resolution  is useful to improve the image quality  (see Section~\ref{sec:spd} for detail). 
The deterioration of the SPD resolution is mainly due to the well-known ambiguity from multi-hits on the orthogonal 2D strip readout in the TPC.
Usually, $n$-hits on anode and cathode strips in the same timing generate $n^2$ hit points  (Figs.~\ref{fig:tracks}a and b),
and thus  a part of the track running horizontally to the $\mu$-PIC is obtained as a square instead of a line. 
In this analysis, a timing resolution of near 10 ns from the clock of the FPGA (100 MHz) was used. 
To address this issue, improvement of the ETCC is being carried out  (Sections~\ref{sec:spd} and \ref{sec:discussion}). 

\section{Performance under Intense Radiation Background}\label{sec:intense_bg}
In order to investigate quantitatively the performance of the ETCC under intense
radiation conditions similar to those found in space, we performed a test using
a 140 MeV proton beam incident on a water target to produce a diffuse background of fast neutrons and MeV gamma-rays.
The 30 cm-cubic ETCC was placed at a distance of 1.3~m from the target,
and direct gammas from the water target were shielded with  lead blocks, allowing us to uniformly irradiate the ETCC \citep{Matsuoka_2015}.
The neutron-gamma ratio and overall intensity can be adjusted by altering the beam intensity and size of the water target.
Details of the radiation condition created by a proton beam on a water target and its simulation will be discussed elsewhere. 
For the present measurement, the ratio of neutrons to gamma was estimated to be  $\sim$1:1 with  both the simulation and
the observed spectra of a neutron monitor located near the water target.
This is similar to the composition of the radiation field observed with  SMILE-I at the balloon altitude at the middle latitude
of the northern hemisphere \citep{Takada_2011a}.

\begin{figure*}[htbp]
\includegraphics[width=\linewidth]{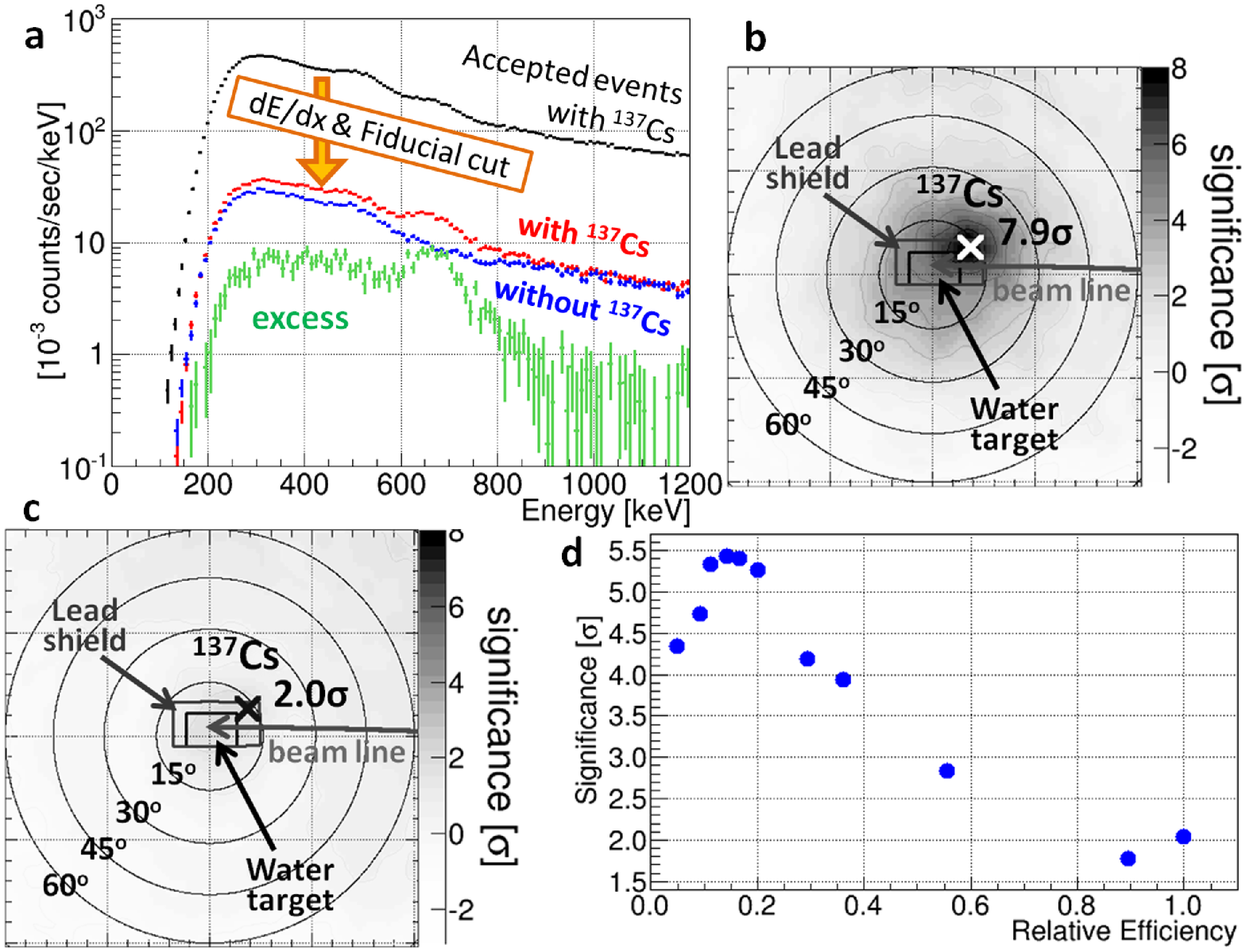}	
\vspace{1ex}
\caption{
(a)  The four energy spectra for reconstructed events under the intense radiation, which is generated by a 140 MeV proton beam, 
before and after applying $\dedx$ and fiducial cuts with a $^{137}$Cs source, without $^{137}$Cs as a background, 
and the spectrum of the excess gammas from $^{137}$Cs, which is derived  by subtracting the background spectrum. 
In addition, (b) is the observed image of $^{137}$Cs (0.8 MBq) set at a 1 m distance from the ETCC using electron tracking. 
(c) is a conventional Compton image without electron track information with the same data of (b). 
(d) Variation of the significance without electron track information as a function of the limiting ratio of
the energy range of an electron in the TPC, where we apply additional cuts on the energy deposit in the TPC,
and the horizontal axis is the ratio of the remained events after this cut to the events with the use of electron track information is applied.
\label{fig:RCNP}
}
\end{figure*}

During the 5 day test, the ETCC was operated stably at the counting rates
up to 1 kHz, or more than 5 times higher than that expected under balloon conditions \citep{Matsuoka_2015}.
Figure~\ref{fig:dedx}b shows the distribution of the measured $\dedx$ (similar to Fig.~\ref{fig:dedx}a), 
where the recoil protons due to scattering of fast neutrons are located at the far right of the plot.
Here, the Compton electrons are clearly separated from the intense backgrounds,
allowing us to reduce the background contribution by one order of magnitude
(see  the energy spectra in  Fig.~\ref{fig:RCNP}a),
using $\dedx$ and fiducial cuts. In this way, we successfully obtained  a clear image of a $^{137}$Cs source (662 keV gammas) set at a distance of 1~m from the ETCC
 (Fig.~\ref{fig:RCNP}b). The image of the source appears at the expected position with a significance of 7.9$\sigma$ (SPD resolution of 200$^\circ$), 
after  gammas with energies in the range 662$\pm$66 keV are selected,
where the  significance was estimated by comparing the spectrum of the $^{137}$Cs source with
the ring-shaped background region between 60$^\circ$ and 90$^\circ$ from the center of the FOV (i.e. the center of the target).
It is also noted that the detection efficiency was consistent with
that measured in the laboratory using the same cuts, as in Fig.~\ref{fig:effective_area}.

For comparison, we performed the image reconstruction using a conventional Compton imaging technique,
including no information about the electron track (Fig.~\ref{fig:RCNP}c).
In this case, only a small enhancement of 2.0$\sigma$ was obtained for the source signal.
The significance could be increased to 5.4$\sigma$ by applying a tight cut on
the electron energy measured in the TPC (Fig.~\ref{fig:RCNP}d). However, such a tight cut would also remove  about 85\% of Compton events,
as compared to our analysis using the detailed information of the electron tracks,
which corresponds  to a decrease in the efficiency to $\sim$15\% of the nominal value (as indicated with  the black square in Fig.~\ref{fig:effective_area}).

This clearly demonstrates  the effectiveness of the ETCC and the additional detailed tracking information
for gamma-ray imaging, as compared to traditional CC.
The requirement of one fully contained electron in the TPC removes the major background sources
of electron escape and reabsorption events.
In addition, the $\dedx$ cut effectively rejects all ``neutrons and cosmic rays.''
Thus,  we are able to select pure Compton-scattering events without loss of efficiency by applying these two filters,
i.e., the fully contained recoil electrons and the $\dedx$ cut (Fig.~\ref{fig:dedx}).
Such a robust event selection also gives the added benefit of enabling us to simulate the fluxes and
images of gammas from both celestial points and extended objects with high reliability. 

\begin{figure*}[htb]
\includegraphics[width=\linewidth]{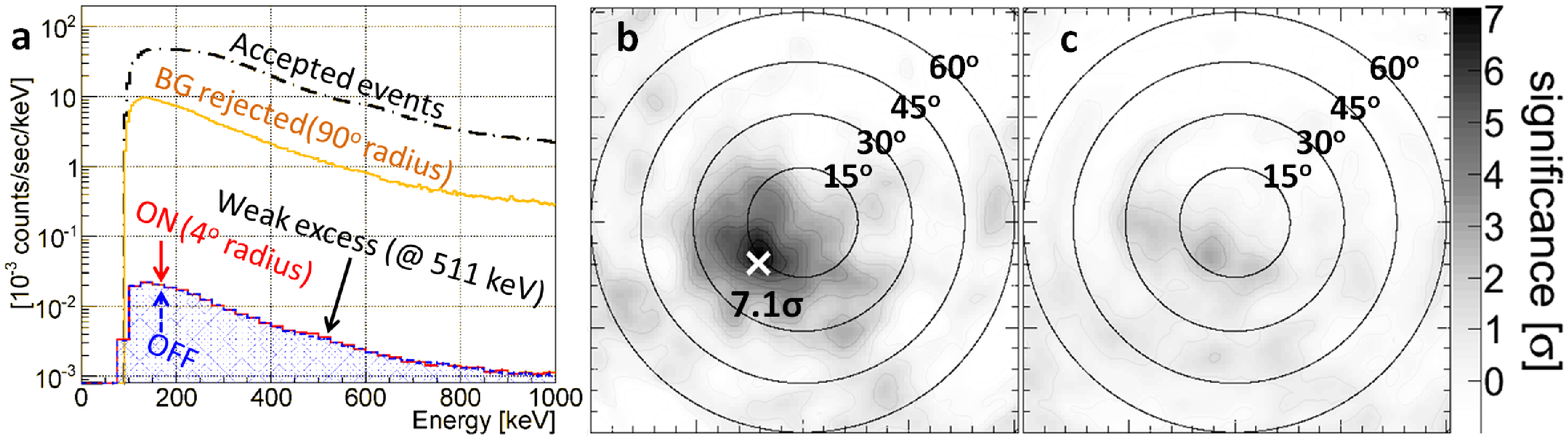}	
\vspace{1ex}
\caption{
(a) Energy spectra of a super-weak point source (27 kBq $^{22}$Na) at a distance of 5.5 m. 
(b) Observed image with 7.1$\sigma$ using electron tracking and SPD resolution of 200$^\circ$ before the improvement, 
and in (c), no significant enhancement ($<$3$\sigma$) appears in the absence of electron tracking.
\label{fig:weak_src}
}
\end{figure*}

In a separate experiment, a measurement was carried out, using a 27 kBq equivalent $^{22}$Na source (511~keV) set at a distance of 5.5~m
from the ETCC, in order to study the performance under poor signal-to-noise ratio (S/N) conditions
similar to those present in  astronomical observations. This setup was expected  to provide a S/N of about
a half of that expected from the observation of  Crab under balloon conditions (expected S/N of $\sim$0.005).
By applying the two cuts explained in  Section~\ref{sec:performance}, we obtained a clear image with a significance of 7.1$\sigma$  in
100~hr observation for an SPD resolution of 200$^\circ$ and gamma energies of 511$\pm$51 keV
 (Fig.~\ref{fig:weak_src}b). Fig.~\ref{fig:weak_src}a gives the  associated energy spectra.  
The error  of the measured fluxes was $\sim$20\%, as estimated from the difference between
the measurement result and that expected from the effective area in Fig.~\ref{fig:effective_area}.
On the other hand, no significant enhancement ($<3\sigma$) was observed in the absence of electron-tracking information,
even after  a tight energy-cut is applied (Fig.~\ref{fig:weak_src}c).
This result of Fig.~\ref{fig:weak_src}b supports a significant  detection of  Crab with at least 5$\sigma$
in the energy range of 200-600 keV in  4~hr of observation, even with  a poor SPD resolution of 200$^\circ$,
where we take  into account by respective factors of 4 and 3  increases in the effective area and 
 in the flux of  Crab in the energy range of 200--400 keV.

\section{Imaging with use of SPD }\label{sec:spd}
Unclean images are  another serious problem in the CC.  However it has been dramatically improved by determining the SPD for all the events. 

\begin{figure*}[htb]
\includegraphics[width=\linewidth]{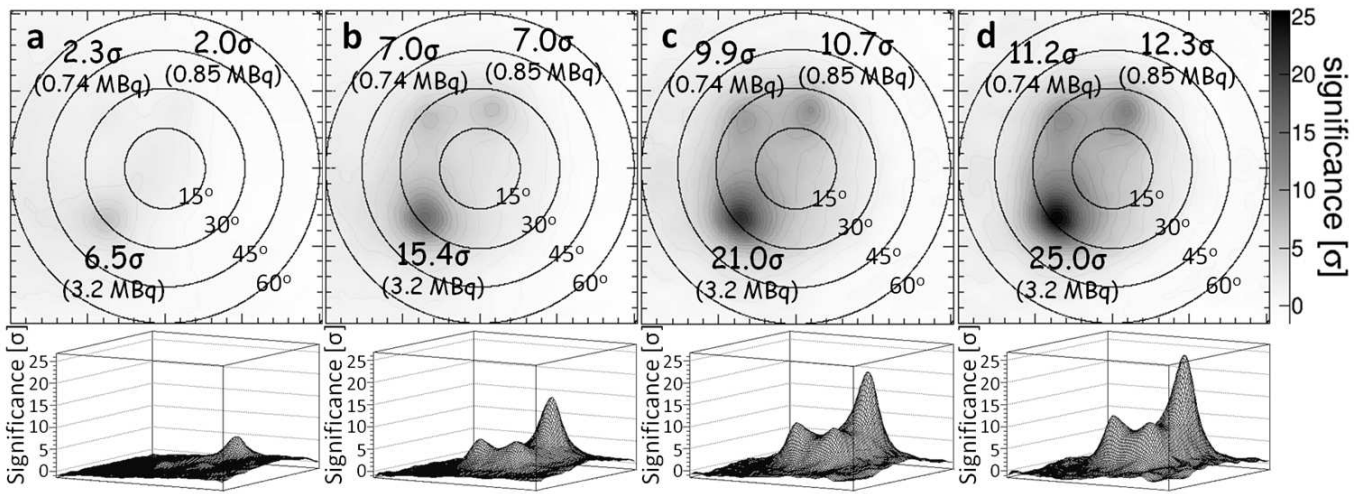}	
\vspace{1ex}
\caption{
Significance maps and contours of three $^{137}$Cs sources with different intensities 
at a distance of 2 m, obtained by (a) accumulating the annulus without tracking information,  
and (b) using the arcs with SPD resolution of 200$^\circ$ before the improvement.   
Same figures after the improvement are (c) and (d) for an SPD resolution of 100$^\circ$ 
and the combination of SPD resolutions of 90$^\circ$ (recoil electron energy range of 10-80 keV) and 45$^\circ$ ($>$80 keV), respectively.
For (d), 
the most reliable SPD resolution that maximizes the imaging significance
was estimated for each energy range (10-80 keV and $>$80 keV), although we should note that the significances are moderately dependent on the SPD resolutions.
These estimated SPD resolutions were better than that in Figs.~\ref{fig:arm_spd}d and e, as mentioned in the caption of Fig.~\ref{fig:arm_spd}. 
\label{fig:3src}
}
\end{figure*}

In CCs, images are obtained by accumulating Compton annuli on the celestial sphere, 
where a fair amount of background event annuli overlap  the target region and produce  serious artifacts. 
Figures~\ref{fig:3src}a and \ref{fig:3src}b show the significance maps of three $^{137}$Cs sources with different intensities, 
put at a distance of 2 m, with respective  annuli and SPD arcs being overlaid.  
The  significances are improved by a factor of $\sim$3,  even with  the worst possible SPD resolution of $\sim$200$^\circ$. 
To obtain these images, the probability functions of one gamma were  defined as an annulus with the area normalized to 1 for a CC \citep{Schonfelder_1993} and
as a  Gaussian  normalized with the SPD resolution for the ETCC.

The  SPD resolution was determined by the uncertainty of the reconstruction method rather than the multiple scattering of a recoil electron (Section~\ref{sec:etcc}). 
We have recently successfully improved the  time resolution of the coincidence between anodes and cathodes  to 1 ns from 10 ns by correcting the time walk of each hit pixel using the TOT. 
 As a result, the reduction of the number of multi hits in a track  provides
a very clear image of the track  (Figs.~\ref{fig:tracks}c and \ref{fig:tracks}d). 
Such a clear track enables us to measure the SPD at a distance of 1 cm from the scattering point,
which reduces the effect of multiple scattering in the gas. 
Figure~\ref{fig:arm_spd}c shows a correlation between  the improved SPD and the electron energy. 
 The spread of the SPD obviously shrinks, as the electron energy increases ($>$80 keV)
 (Figs.~\ref{fig:arm_spd}d and \ref{fig:arm_spd}e).
The SPD resolutions are nearly consistent with the multiple scattering angles in Ar gas. 
Next, we reanalyzed the case presented in  Fig.~\ref{fig:3src}b with  a SPD resolution of 100$^\circ$ and two SPD resolutions optimized for low energy electrons ($<$80 keV)
and high energy ones ($>$80 keV), and plotted the results in Figures~\ref{fig:3src}c and \ref{fig:3src}d, respectively. 
We found a massive  increase in the signal by  $\sim$10$\sigma$. 
In Fig.~\ref{fig:3src}d, 
the most reliable SPD resolution that  maximizes the imaging significance
was estimated for each energy range (10-80 keV and $>$80 keV), although we should note that the significances are moderately dependent on the SPD resolutions.
These value are better than the SPD resolutions in Figs.~\ref{fig:arm_spd}d and \ref{fig:arm_spd}e (see  the caption of Fig.~\ref{fig:arm_spd}).

\section{Discussion}\label{sec:discussion}
 Our ETCC provides
a robust recipe for both removing the huge backgrounds and producing clear images,
and as a result, gives nearly  10 times better overall sensitivity than that of conventional CC.
The next important step would be  the improvement of the SPD resolution up to the multiple scattering-angle 
in order to exploit the potential  of electron tracking.

\begin{figure*}[htb]
\includegraphics[width=\linewidth]{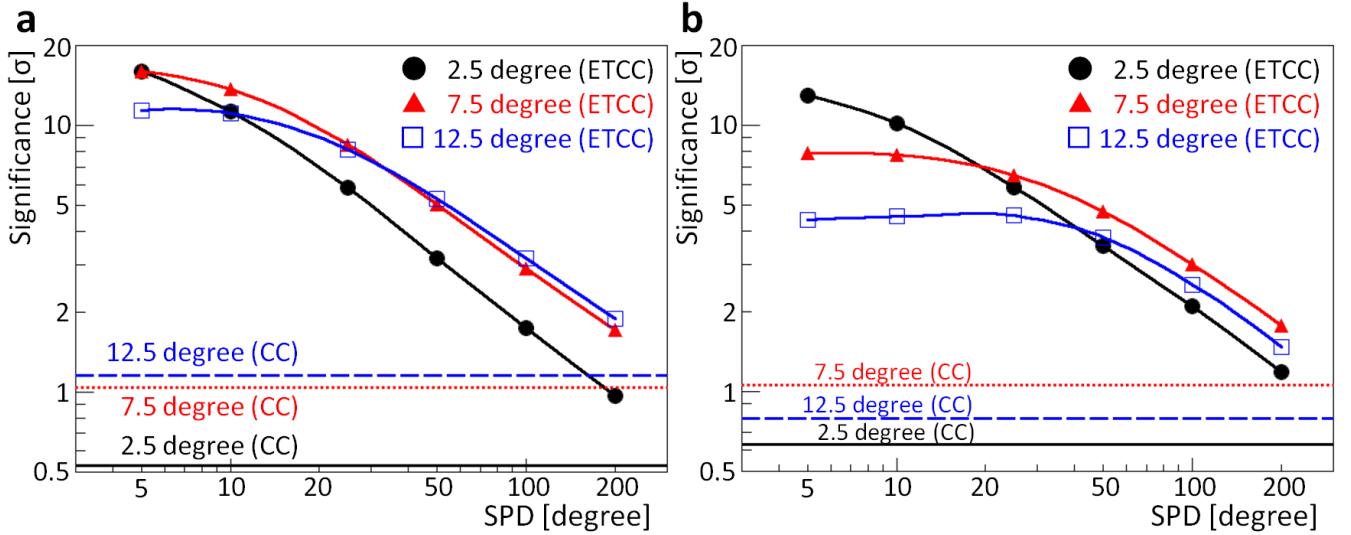}	
\vspace{1ex}
\caption{
Simulated significances as a function of the SPD resolution (FWHM) for the signal to noise ratio of 10$^{-3}$, 
using 10$^3$ signal events and 10$^6$ background events randomly distributed on the 2$\pi$ str, 
where three accumulated regions on the FOV are indicated with angular radii of 2.5$^\circ$ (circle), 7.5$^\circ$ (triangle) and 12.5$^\circ$ (box). 
 The ARM resolutions of 2$^\circ$ and 5$^\circ$ (FWHM) are assumed for (a) and (b) respectively, 
and the Compton scattering angle $\varphi $ in Fig.~\ref{fig:etcc_schematic}a is restricted within 60$^\circ$.
\label{fig:sigma_spd}
}
\end{figure*}

It is well known that  the annulus of conventional CC leads to multiple intersections and a wide spread of the PSF.
A better SPD resolution is naively expected to increase the significance by a factor roughly proportional to 180$^\circ$/(SPD resolution).
Our simulation results appear to confirm it Figures~\ref{fig:sigma_spd}a and \ref{fig:sigma_spd}b, 
where a S/N of $1:10^3$ (or $10^3 :10^6$ total events) is  assumed.
In the simulation, the variation of the significance as a function of the SPD resolution was calculated 
for three accumulated regions on the FOV and two ARM resolutions,
while constraining the Compton scattering angle $\varphi $ (Fig.~\ref{fig:etcc_schematic}a) to smaller  than 60$^\circ$.

\begin{figure}[htb]
\includegraphics[width=\linewidth]{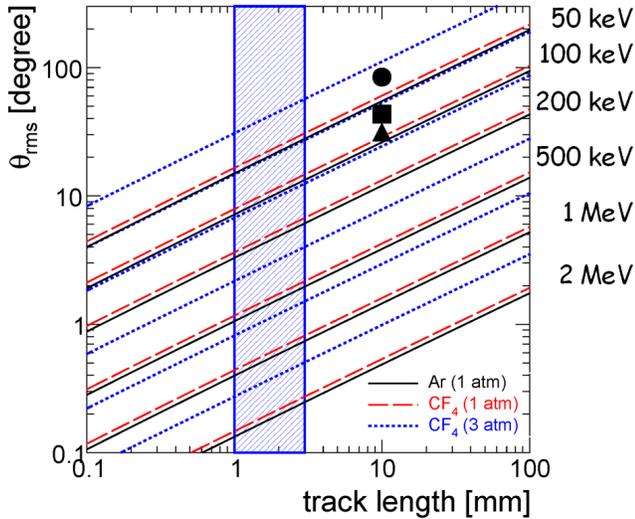}	
\vspace{1ex}
\caption{
Dependence of the multiple-scattering angles of a recoil electron along  the length of the track 
from the scattering point in Ar and CF$_4$ gases with varying electron energies,  
where the normal pressure is assumed for both gases.
The circle, square and triangle are the root mean square of SPD resolutions of 200$^\circ$ (before improvement), 
90$^\circ$ and 45$^\circ$ (from Fig.~\ref{fig:3src}d), respectively. 
The hatched region corresponds to the SPD resolutions expected by fine track sampling of 400 $\mu$m pitch with a fine fitting.
\label{fig:multiple_scatter}
}
\end{figure}

Figure~\ref{fig:multiple_scatter} shows the calculated dependence of the multiple scattering 
angle of the recoil electron along  the length of the track used for the angle measurement for Ar and CF$_4$ gases at the normal pressure and CF$_4$ at 3 atm, where
the  calculation was performed with  GEANT4, in which we employed  an empirical expression for the multiple scattering based on experimental data \citep{Attwood_2006}.
 We expect to improve the SPD resolution to the hatched region in this figure (see the discussion later in this section).
Thus, the SPD resolutions for gammas above 500 keV (with recoil-electron energy of $>$200 keV) and above 1~MeV will be reduced to within 20$^\circ$ and 10$^\circ$, respectively.

\begin{figure}[htb]
\includegraphics[width=\linewidth]{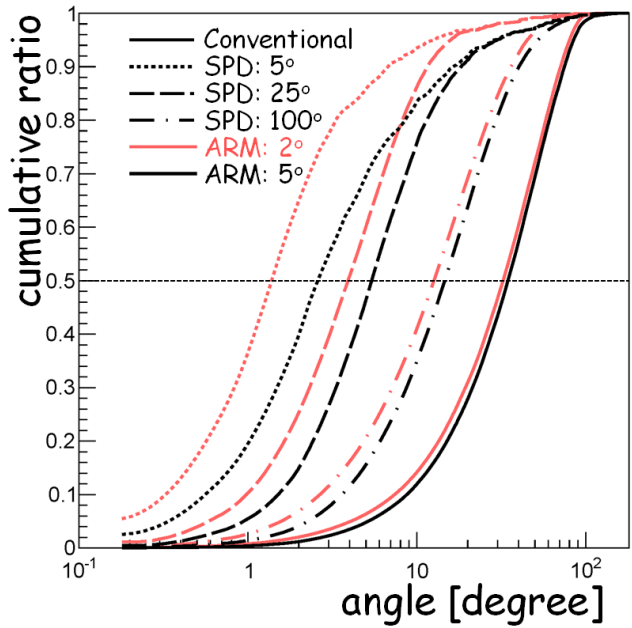}	
\vspace{1ex}
\caption{
Cumulative ratio in the PSF for gammas from a  point source as a function of its angular radius
for several SPD resolution and two ARMS; pastel red and black lines are the ARM resolutions of 2$^\circ$ and 5$^\circ$, respectively, 
and solid, dotted, dashed and dotted-dashed lines are conventional CC, SPD resolutions of 5$^\circ$, 25$^\circ$, and 100$^\circ$, respectively.
\label{fig:psf}
}
\end{figure}

The cumulative ratio of the PSF for gammas emitted from a point source is plotted as a function of its angular radius
for three  SPD resolutions (100$^\circ$, 25$^\circ$, and 5$^\circ$) in Fig.~\ref{fig:psf}, along with the annulus of a conventional CC, where
 two ARM resolutions of 2$^\circ$ and 5$^\circ$ (FWHM) are  assumed.
Hereafter, PSF ($\theta$) is defined to contain a half of the gammas emitted from the point source within the angular radius $\theta$.
First, from Fig.~\ref{fig:psf}, we note that the PSF is predominantly dependent on the worse one between  the SPD and  ARM resolutions,
where an SPD resolution of $\sim$20$^\circ$ gives a similar dispersion as an ARM resolution of 5$^\circ$
when projected on the celestial sphere.
  The improvement in the significance due to better ARM resolutions is only seen for
the case with the  SPD resolution of better than $\sim$30$^\circ$  (Fig.~\ref{fig:psf}).
Thus, the improvement of the ARM resolution from $5^\circ$ to $2^\circ$ is  effective in  improving the PSF only if the SPD resolution is also improved below $\sim$30$^\circ$.
The PSF for conventional CCs, on the other hand, appears to be determined by the Compton scattering angle $\varphi$ of Fig.~\ref{fig:etcc_schematic}a, 
judging from the fact that  the cumulative ratio of the PSF reaches  1 at $\sim$120$^\circ$, or the largest diameter of the scattering $\varphi$.
Thus, only an improved SPD resolution can counteract the spread of signals due to the annulus in a CC,
and can recover a point-like PSF similar to telescopes in other energy bands.

The wide spread of the PSF for poor SPD resolutions and conventional CCs
is due to the normalization of the probability on the arc and annulus  (Section~\ref{sec:spd}), 
which is necessary to keep the background level statistically similar for any SPD resolution, including an annulus.
 Maximum Likelihood Expectation Maximization (ML-EM) has been used in CCs to compensate for the spread
and sharpen the PSF of gamma-ray sources from the background \citep{Schonfelder_1993, Wilderman_1998, Bandstra_2011}. 
When the statistics of the signals are ideal, ML-EM can extract the maximum signal as determined by the effective area and PSF (ARM resolution).
However, even with ML-EM, it is difficult to reproduce statistically poor signals from a huge background with the  uncertainty at the similar level as  the signal.
We also note that an inherent  risk of introducing artifacts for such background-dominated data should not be underestimated, particularly in  MeV gamma-ray astronomy.
Thus, conventional CCs have looked to detect signals for bright celestial objects only, for which the ML-EM works optimally.
However,  moderate-intensity objects, whose flux was judged to be  sufficient to allow detection, based on  the sensitivity estimated  from the effective area and PSF,
 actually turned out to be difficult to detect due to the breakdown of the ML-EM method.
On the other hand, ML-EM would function well for an ETCC, thanks  to its sharp PSF and low background. 
More fundamentally, the sensitivity would be determined solely by the effective area and the PSF,
and supplemental tools, such as ML-EM, only serve to improve the significance by several factors by applying some constraints. 

Considering all those,  how would a better SPD resolution be attained, using the present technology of the ETCC?
In the present detector technology, a gaseous TPC with a micro-pattern detector provides very accurate 3D tracking for electrons in the energy range from keV to MeV.
Even a high-resolution CCD with a 10 $\mu$m pixel pitch and 2D tracking has been reported to provide
an SPD resolution of only $\sim$200$^\circ$ FWHM for $\sim$100 keV for 2D tracking \citep{Vetter_2011}.
We have just begun the detailed study of the 3D reconstruction of recoil electrons in the gas of the ETCC, 
and  expect to be able to improve the uncertainty of the Compton-scattering point from the current $\sim$8 mm down to a few millimeters.
This will improve both the SPD and ARM resolutions to about a half of the present value and  to the calculated value in Fig.~\ref{fig:tracks}, respectively. 

The selection of the gas for the TPC is the most important factor for  the performance of the ETCC.
In general, a scatterer of CCs should provide a large cross-section of Compton scattering,
while also having  a small cross-section for photo absorption to minimize the  background.
In addition, a small diffusion constant and small multiple scattering for the recoil electron are very important for the ETCC.
 All of the  requirements above can be met with  CF$_4$ gas in the ETCC.
Then,  we are planning to use a 3 atm CF$_4$ gas in the next version of the ETCC.
CF$_4$ gas has a large Compton-scattering cross-section due to the large number of electrons per molecule (44),
while simultaneously suppressing the photo-absorption cross-section, which is proportional to $Z^5$
in the Born approximation (with the atomic  number $Z$ = 9 for F).
Its  diffusion constant is smaller than that of  Ar gas.  Hence, it is possible to develop  a larger TPC with a drift length greater than 50~cm,
where a positional error at the MPGD due to the fluctuation of the centroid of the diffusive drift-electron cloud
over a 50-cm drift length is estimated to be $\sim$150~$\mu$m in a 3 atm CF$_4$ gas.

Furthermore, with the use of CF$_4$,  the SPD resolution, which is
the most important parameter for the ETCC, is expected to be improved in the following reason. 
For pressurized gas, the multiple scattering angle is proportional to the product of $\surd$(pressure) and $\surd$(track range).
Whereas  the multiple scattering angle per length of CF$_4$ is similar to that of Ar gas  (Fig.~\ref{fig:multiple_scatter}),
$\dedx$ is 3 times larger, and the  diffusion constant is smaller, than those of Ar gas. Then, combined with a smaller track sampling of 400 $\mu$m (reduced from the present 800 $\mu$m),
it would significantly improve the fitting resolution of the track.
With this expected improvement, it would be possible  to reduce the track range needed to measure  the SPD from 1 cm down to a few mm,
which would exceed the increase of the multiple scattering due to the increase in gas pressure to 3 atm.
Thus, an ETCC with pressured CF$_4$ gas would give  a slightly better PSF than that of Ar gas at the normal pressure.
This fact, combined with the properties described above, makes CF$_4$ gas extremely well suited to provide the optimum performance for our ETCC.

Next, although the introduction of $\dedx$ and SPD successfully resolved the two problems of huge backgrounds and unclear images ,
we should also consider the backgrounds that cannot be removed in theory  with  $\dedx$ and/or SPD, i.e.,  
the simultaneous emission of X- and gamma-rays or two gamma-rays from  radioisotopes generated by cosmic rays, 
and accidental coincidences between the TPC and PSAs.
Due to the large cross-section for photo absorption of X-rays,
false Compton events arise from a photo-electric absorption of one gamma-ray (or X-ray) in the TPC and a random coincidence of a second gamma-ray in the PSA.
Indeed, those false events  were reported to be  one of the dominant backgrounds in  COMPTEL \citep{Weidenspointner_2001}.
However, the small photo-electric absorption cross-section of CF$_4$ ($\sim$1/5 of that of Ar with the same number density of atoms)
makes it considerably less sensitive to  X-rays compared to Ar and Si.
Absorption probabilities for X-rays at an energy of a few tens of keV  in a 50 cm-cubic 3 atm CF$_4$ and 8 atm Ar
are transparent ($\sim$10\% at 30~keV) and opaque ($>$80\%), respectively, where the respective pressures are chosen so that they have the same Compton-scattering probability.
Such insensitivity to X-rays of CF$_4$ also reduces the false Compton events, because  most of the single hits in the TPC are due to X-rays.
Furthermore, the ETCC provides another tool to remove such backgrounds via a kinematical test using the angle $\alpha$ of Fig.~\ref{fig:etcc_schematic}a,
which was introduced as a new parameter for the ETCC in SMILE-I  (Section~\ref{sec:etcc}).
 The poor angular resolution for the direction of the recoil electron would result in the rejection of a number of good Compton events.
For that reason, we have not used the kinematic test for SMILE-II.
However, because such events do not satisfy Compton kinematics, 
the angle $\alpha$ should be randomly distributed between $-90^\circ$ and $90^\circ$.
Further improvement of the directional resolution of the recoil electron will reduce the uncertainty of $\alpha$ to $<$20$^\circ$ (from the current $\sim$40$^\circ$ at FWHM),
which will allow us to suppress such background by nearly one order of magnitude.

A final remark is that  we have developed the ETCC,  using well-established technologies (i.e., gas counters and scintillators),
and even a 4-module satellite ETCC would need  only $\sim$10$^4$ read-out channels for TPC and PSAs with low power electronics and no specific cooling system.
In addition,  due to the strong background rejection capability, no active shield would be  required.
Then, the component with the largest contribution to the weight would be  the scintillators (more than half the total weight),
which is intrinsically needed to absorb and detect gamma rays.
Thus, the ETCC would be  the lightest CC, and would provide  the lowest background-radiation level in space.

\section{Conclusion}
\begin{figure}[htb]
\includegraphics[width=\linewidth]{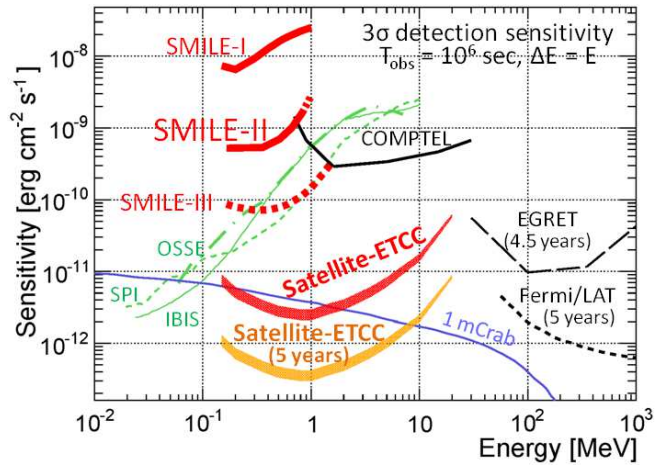}	
\vspace{1ex}
\caption{
Sensitivities with 3$\sigma$ detection of SMILE-II, III and Satellite-ETCC. 
 Green and black lines are from previous results \citep{Atwood_2009, Takahashi_2013}.
\label{fig:sensitivity}
}
\end{figure}

We have revealed that the SPD dramatically recovers the gammas detected in the effective area
within several degrees of the target position, 
whereas they  exude sparsely over the FOV in conventional CC.
With this,  $\sim$10 times better sensitivity than CC per unit  effective area will be certainly expected (Fig.~\ref{fig:sigma_spd}). 
Also, the radical reduction of almost all the instrumental background is found to be attainable by the use of the measured $\dedx$ of the recoil electron, 
a kinematical test of the $\alpha$ angle and the optimization of the scattering material such as CF$_4$.
Our ETCC has changed delicate CCs to a very tough and reliable gamma imaging device under intense radiation conditions, notably the  space. 
Thus, by resolving two serious problems of the MeV gamma-ray astronomical observations, 
we have reached the intrinsic sensitivity determined by its effective area, PSF, and the background of cosmic diffuse gammas, as follows.
Figure~\ref{fig:sensitivity} shows the expected sensitivities of the SMILE balloon experiments, 
along with a satellite-ETCC consisting of four 50 cm-cubic ETCCs, which, with its large FoV, 
provides the sensitivity of $\sim$2$\times$10$^{-12}$~erg~cm$^{-2}$~s$^{-1}$ at 1 MeV for 10$^6$~s observation, 
and $\sim$3$\times$10$^{-13}$~erg~cm$^{-2}$~s$^{-1}$ for continuum gammas
with a sensitivity of $\sim$10$^{-7}$ gamma cm$^{-2}$~s$^{-1}$ for line gamma-rays
in  5 years operation ($>$100 times better than those of COMPTEL in  9 years), where 
a duty factor of 0.5 in the operation is assumed.
Note that the  green and black lines are from the previous results \citep{Atwood_2009, Takahashi_2013}.
  The background
observed in SMILE-I was used for the SMILE-II and SMILE-III simulations. 
For the background for the Satellite-ETCC, we used the value twice as  the extragalactic diffuse gamma flux in 0.1--5 MeV reported with  SMM \citep{Watanabe_1999} and COMPTEL \citep{Weidenspointner_2000}, 
assuming the instrumental background to be at the same level as the extragalactic diffuse gamma. 
The PSF ($\theta$ = 1.2$^\circ$) with the ARM and SPD resolutions of 2$^\circ$ and 5$^\circ$ was used for the Satellite-ETCC, 
and the PSF (4$^\circ$) with the ARM and SPD resolutions of 5$^\circ$ and 25$^\circ$ was used for the SMILE-II and III.
 One half of the detected gamma-rays from a  point source were used in  the calculations of the sensitivity.
The application of ML-EM to the ETCCs may improve the sensitivities by several factors. 
However, it is not essential because we successfully provide, for the first time,
 quantitatively reliable sensitivities for CCs without  the use of an optimization algorithm like the ML-EM method. 
Thus, satellite-ETCC would access the deepest universe in high energy astronomy.

Furthermore, with its excellent sensitivity it would detect  $\sim$15 SNe Ia/year at distances up to 60 Mpc
with required statistical accuracy  (Section~\ref{sec:introduction}).
For the distances up to 100 Mpc,  $\sim$500 and nearly a thousand SNe/5 years  ($\sim$5$\sigma$ in  150 days)
 would be 
expected with and without optical coincidence, respectively,  with no observational bias, thanks to the  large FOV and high transparency of MeV gammas.
Note that this sample should include some collapsars, in which production of $^{56}$Ni is considered to be less than that in SN Ia.

Also, the well-defined reconstruction of Compton events with low background  endows the ETCC with an excellent polarimetry 
(simulated modulation factors of 0.6 and 0.5 at 200 and 500 keV, respectively). 
 The Satellite-ETCC would measure $\sim$10\% polarization (3$\sigma$) for 20 mCrab. 
Thus, such  observations would open a new era of astronomy with distinctive detections of thousands of MeV objects.

\acknowledgments
We thank very much  Prof. Keiichi Maeda at Kyoto University and Dr. Alexander Summa at Wurzburg University for the fruitful discussions about SNe.
This study was supported by 
the Japan Society for the Promotion of Science (JSPS) Grant-in-Aid for Scientific Research (S) (21224005),
(A) (20244026), JSPS Grant-in-Aid for Challenging Exploratory Research (23654067, 25610042),
a Grant-in-Aid from the Global COE program ``Next Generation Physics, Spun from Universality and Emergence'' 
from the Ministry of Education, Culture, Sports, Science and Technology (MEXT) of Japan,
and Grant-in-Aid for JSPS Fellows (09J01029, 11J00606, 13J01213).
Also this study was supported by the joint research program of the Solar-Terrestrial Environment Laboratory, Nagoya University  
and the National Institute of Polar Research through General Collaboration Projects no 23-3. 
Some of the electronics development was supported by KEK-DTP and Open-It Consortium.
Finally, we thank the anonymous referee for carefully reading the manuscript and
providing helpful comments to improve it.

\end{document}